\documentclass{kluwer}
\usepackage{graphicx}

\begin{document}
\begin{article}

%%%%%%%%%%%%%%%%%%%%%%%%%%%%%%%%%%%%%%%%%%%%%%%%%%%%%%%%%%%%%%%%%%%%%%%%%%%%%

\begin{opening}

\title{Chemodynamical Modelling of Galaxy Formation and Evolution}

\author{Peter \surname{Berczik} \email{berczik@mao.kiev.ua}}
\institute{
$^1$Main Astronomical Observatory of Ukrainian National Academy of Sciences, \\ 
    Zabolotnoho Str. 27, Kiev, Ukraine \\ 
$^2$Astronomisches Rechen-Institut, Monchhofstrasse 12-14, Heidelberg, Germany}

\author{Gerhard \surname{Hensler}}
\author{Christian \surname{Theis}}
\institute{Institut f\"ur Theoretische Physik und Astrophysik,
University of Kiel, \\ 
Olshausenstr. 40, Kiel, Germany}

\author{Rainer \surname{Spurzem}}
\institute{Astronomisches Rechen-Institut, Monchhofstrasse 12-14, Heidelberg, Germany}

\runningauthor{Peter Berczik et al.}
\runningtitle{Chemodynamical Modelling of Galaxy Formation and Evolution}

%\date{\today}
\date{November 28, 2001}

%%%%%%%%%%%%%%%%%%%%%%%%%%%%%%%%%%%%%%%%%%%%%%%%%%%%%%%%%%%%%%%%%%%%%%%%%%%%%

\begin{abstract}

We present our recently developed 3-dimensional chemodynamical
code for galaxy evolution. This code follows the evolution of
different galactic components like stars, dark matter and
different components of the interstellar medium (ISM), i.e. a
diffuse gaseous phase and the molecular clouds. Stars and dark
matter are treated as collisionless N-body systems. The ISM is
numerically described by a smoothed particle hydrodynamics (SPH)
approach for the diffuse gas and a sticky particle scheme for the
molecular clouds. Additionally, the galactic components are
coupled by several phase transitions like star formation, stellar
death or condensation and evaporation processes within the ISM.

As an example we show the dynamical and chemical evolution of a
star forming dwarf galaxy with a total baryonic mass of $2 \cdot
10^9 M_\odot$. After a moderate collapse phase the stars and the
molecular clouds follow an exponential radial distribution,
whereas the diffuse gas shows a central depression as a result of
stellar feedback. The metallicities of the galactic components
behave quite differently with respect to their temporal evolution
as well as their radial distribution. Especially, the ISM is at no
stage well mixed.

\end{abstract}


%%%%%%%%%%%%%%%%%%%%%%%%%%%%%%%%%%%%%%%%%%%%%%%%%%%%%%%%%%%%%%%%%%%%%%%%%%%%%

\end{opening}

%%%%%%%%%%%%%%%%%%%%%%%%%%%%%%%%%%%%%%%%%%%%%%%%%%%%%%%%%%%%%%%%%%%%%%%%%%%%%

\vskip -0.5cm
\section{The 3d chemodynamical code and first results}

Since several years smoothed particle hydrodynamics (SPH)
calculations have been applied successfully to study the formation
and evolution of galaxies. Its Lagrangian nature as well as its
easy implementation together with standard N-body codes allows for
a simultaneous description of complex dark matter-gas-stellar
systems (e.g. \opencite{NW93}; \opencite{MH96}). Here we present
simulations based on our 3d chemodynamical code. This code
includes many complex effects such as a multi-phase ISM,
cloud-cloud collisions, a drag force between different ISM
components, condensation and evaporation of clouds (CE), star
formation (SF) and a stellar feedback (FB). This code is a further
development of our single phase galactic evolutionary program
\cite{Ber99,Ber2000} including now different gaseous phases.

%---------------------------------------------------------------------------%
\begin{figure}[t]
\tabcapfont
\centerline{%
\begin{tabular}{c@{\hspace{0.1in}}c}
\includegraphics[width=2.2in]{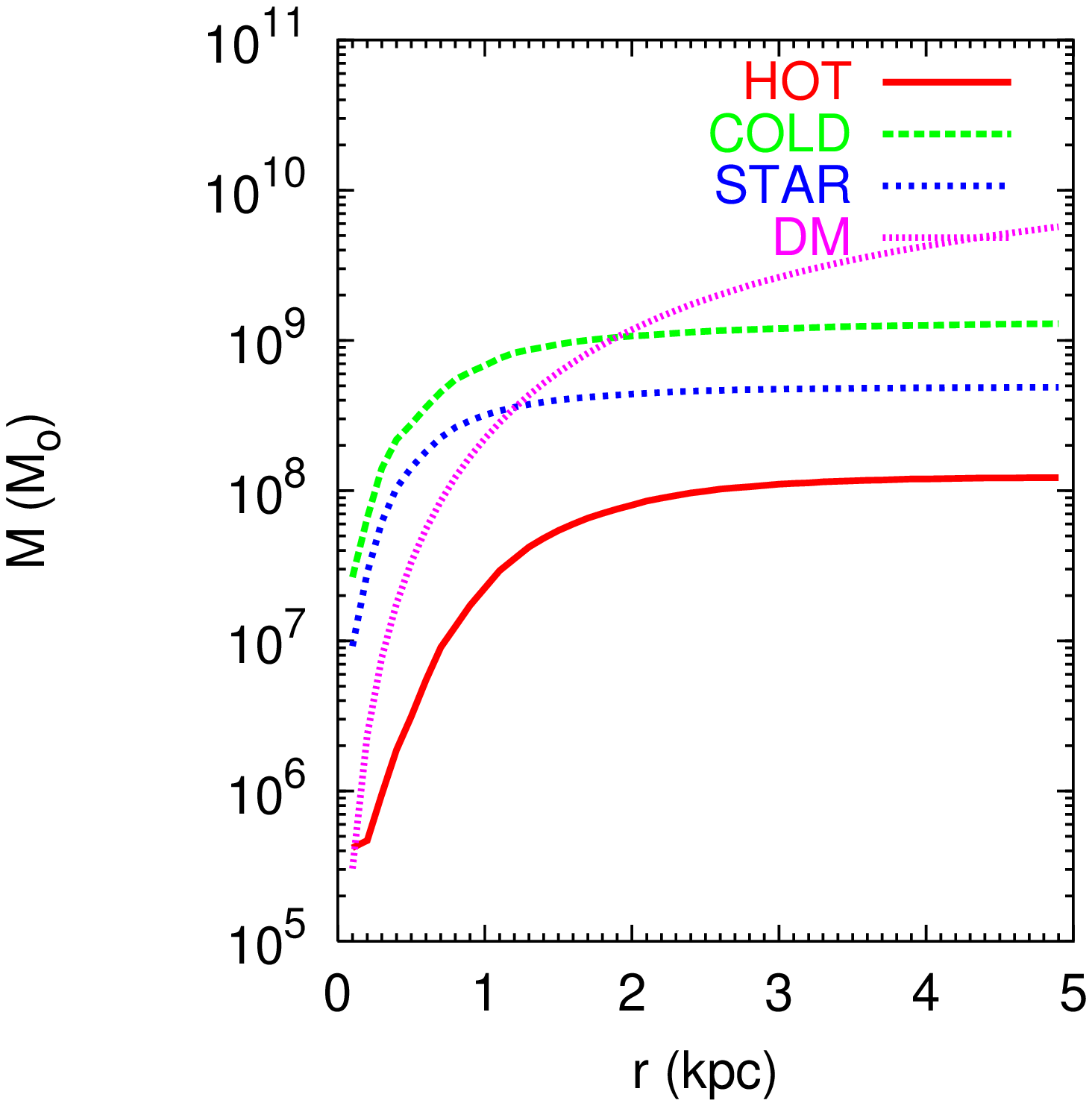} &
\includegraphics[width=2.2in]{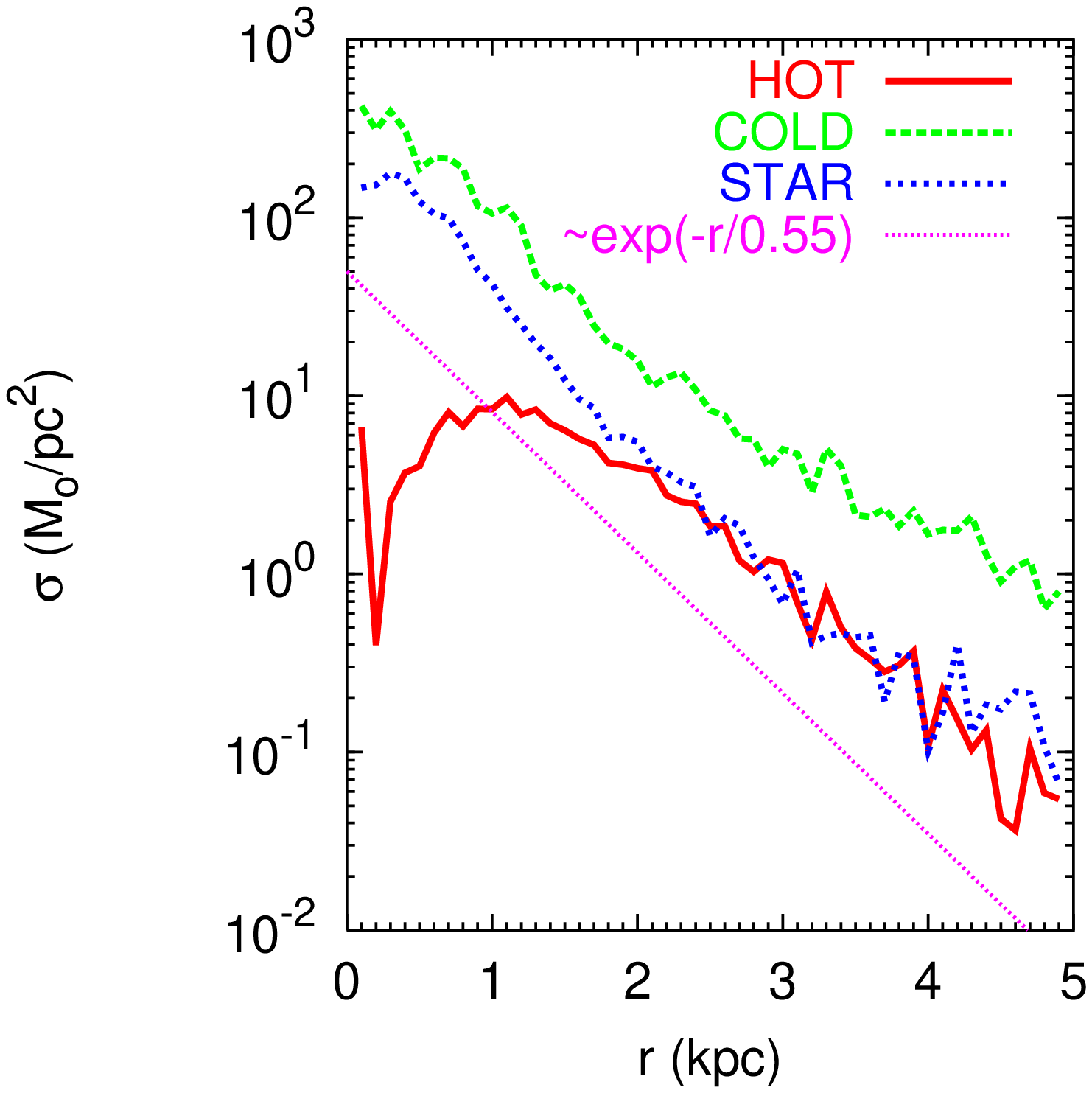}
\end{tabular}}
\caption{The radial distribution of the cumulative mass (left) and the
surface density (right) for the different components in the central
region of the model galaxy after 1~Gyr.}
\label{100-mass&100-sigma}
\end{figure}
%---------------------------------------------------------------------------%

In our new (multi-phase gas) code we use a two component gas
description of the ISM (cold ``clouds'' and ``smooth'' hot SPH phase)
\cite{TBH92,SHT97}. The basic idea is to add a cold cloudy
component to the smooth and hot gas (10$^4$ - 10$^7$ K) described
by SPH. The cold clumps are modeled as N-body particles with some
``viscosity'' \cite{TH93} (cloud-cloud collisions and drag force
between clouds and hot gas component). The cloudy component
interacts with the surrounding hot gas also via condensation and
evaporation processes \cite{CMcKO81,KTH98}. In the code we
introduce also star formation. The ``stellar'' particles are
treated as a dynamically separate N-body component. The cloud
component forms stars. These stars return chemically enriched gas
material and energy to both gaseous phases.

As a test of our new code, we calculate the evolution of an
isolated star forming dwarf galaxy. The initial total gas content
of our dwarf galaxy is $2 \cdot 10^9$ M$_\odot$ (80 \% ``COLD'' + 20
\% ``HOT'') which is placed inside a fixed dark matter halo with
parameters r$_0$ = 2~kpc and $\rho_0 = 0.075$ M$_\odot$/pc$^3$,
using density profile of \opencite{B95}. 
With these parameters the dark matter mass inside the
initial distribution of gas (20~kpc) is $ \approx 2 \cdot 10^{10}$
M$_\odot$. For the initial gas distribution we use a
Plummer-Kuzmin disk with parameters a = 0.1~kpc and b = 2~kpc
\cite{MN75}. The gas initially rotates in centrifugal equilibrium
around the z-axis. We choose the dwarf galaxy as an appropriate
object for our code, because in this case even with a relatively
``small'' number of cold ``clouds'' ($\sim$ 10$^4$) we achieve the
required physical resolution for a realistic description of
individual molecular clouds ($\sim$ 10$^5$ M$_\odot$) as a
separate ``COLD'' particle. In the simulation we use N$_{\rm hot}$ =
10$^4$ SPH and N$_{\rm cold}$ = 10$^4$ ``COLD'' particles. After
1~Gyr more then $10^4$ additional stellar particles are created.

%---------------------------------------------------------------------------%
\begin{figure}[t]
\tabcapfont
\centerline{%
\begin{tabular}{c@{\hspace{0.1in}}c}
\includegraphics[width=2.2in]{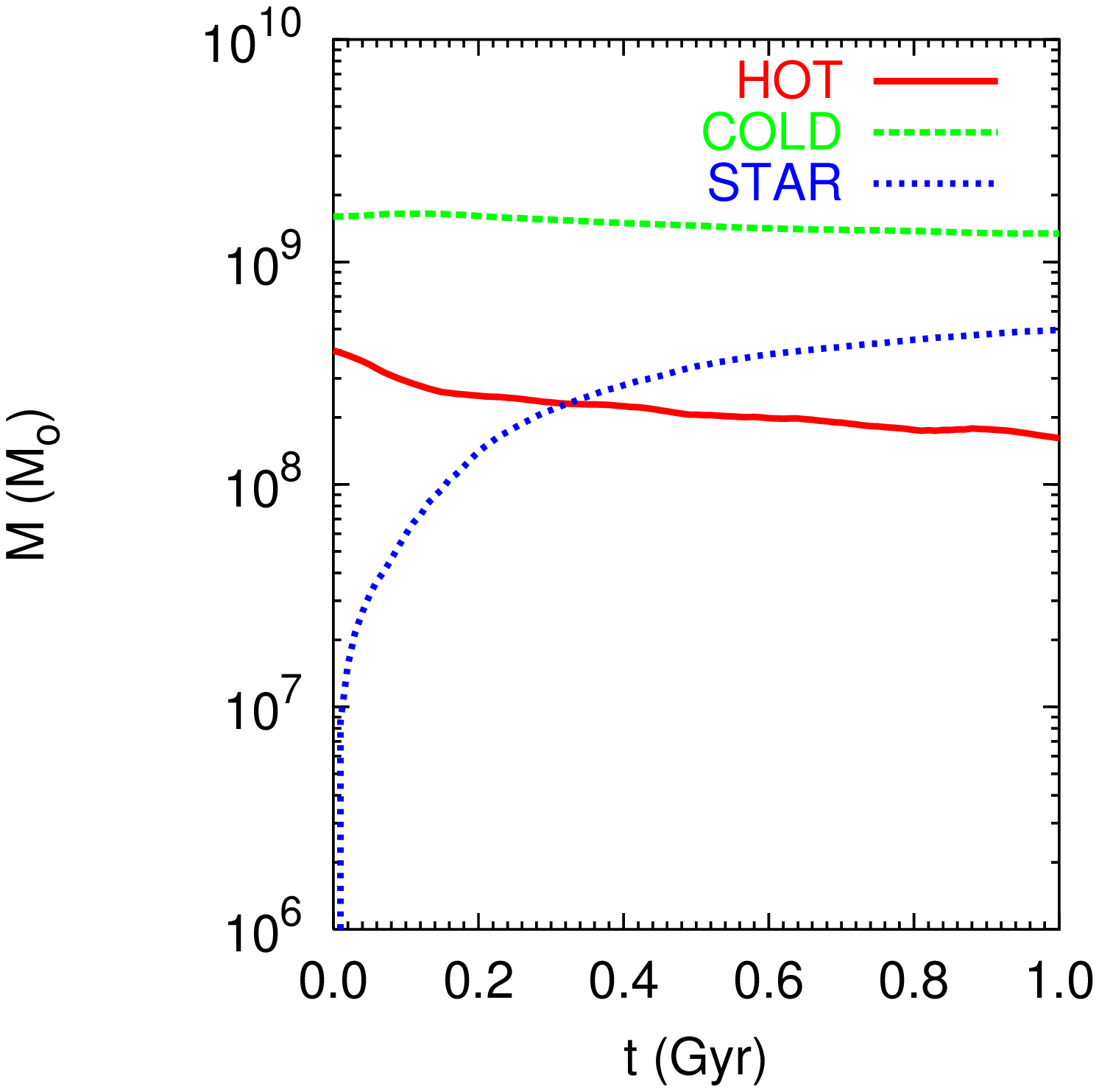} &
\includegraphics[width=2.2in]{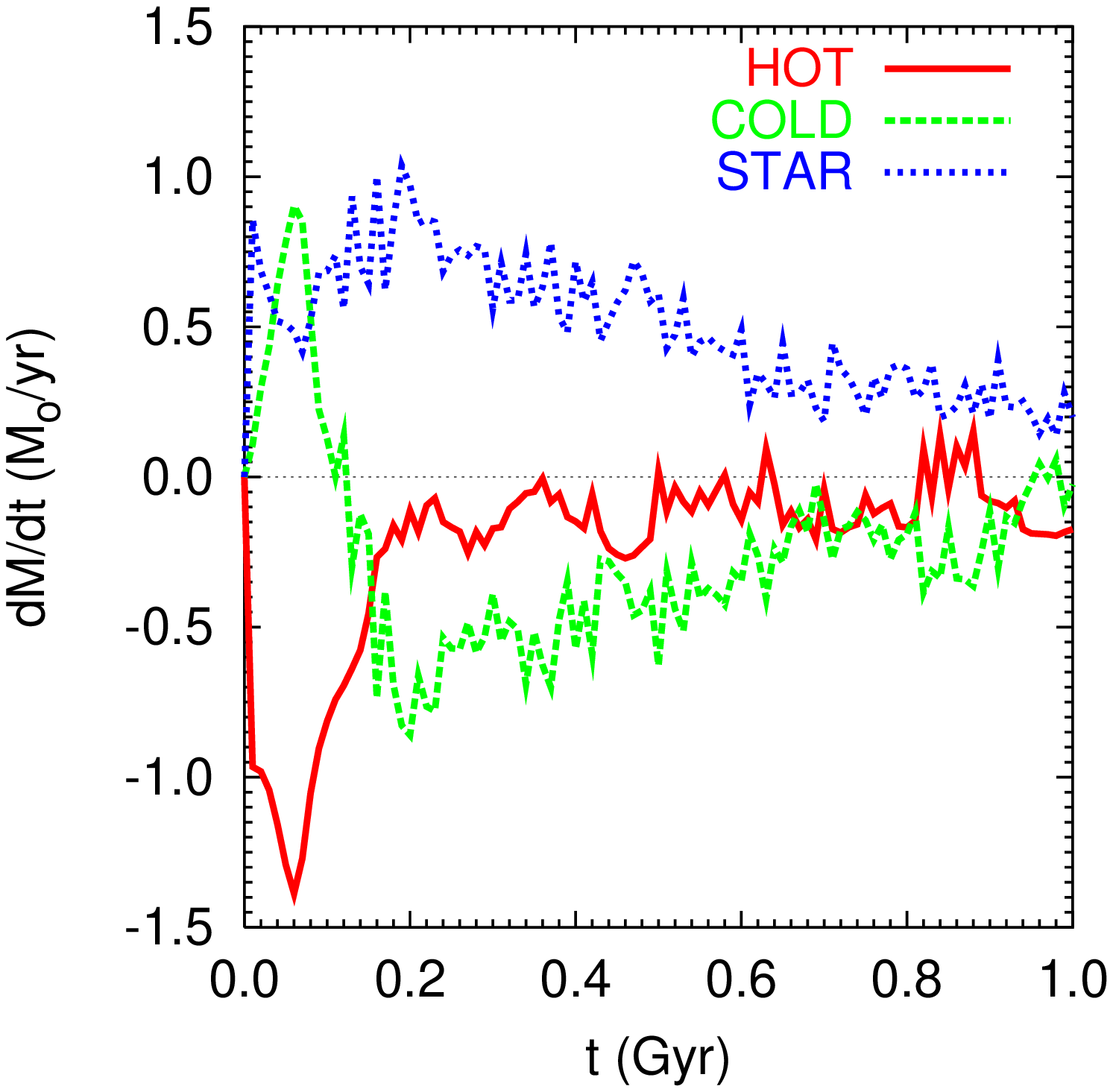}
\end{tabular}}
\caption{The temporal evolution of the mass (left) and mass exchange
rate (right) for the different components of the model galaxy.}
\label{mass-t&dmdt-t}
\end{figure}
%---------------------------------------------------------------------------%

In Fig.~\ref{100-mass&100-sigma} we present the mass and surface
density distribution of the different components in the central
region of the model after 1~Gyr of evolution. In the region up to
$\approx$ 2~kpc the baryonic matter dominates over the DM. The
surface density of the stars can be well approximated by an
exponential disk with a scale length of 0.55~kpc. In the
distribution of hot gas, we see a central ``hole'' ($\approx$
1~kpc), as a result of gas blow-out from the center mainly due to
SNII explosions.

In Fig.~\ref{mass-t&dmdt-t} we present the evolution of the mass
and the mass exchange rate of the different components. The SFR
(i.e. dM$_{\rm STAR}$/dt) peaks to a value of 1 M$_\odot$yr$^{-1}$
after 200 Myrs. Afterwards it drops down to 0.2 M$_\odot$yr$^{-1}$
within several hundred Myrs. Another interesting feature is the
behaviour of the hot gas phase mass exchange. After the initial
violent phase of condensation an equilibrium is established which
gives a hot gas fraction of about 10\% of the total gas mass.

\begin{figure}[t]
\tabcapfont
\centerline{%
\begin{tabular}{c@{\hspace{0.1in}}c}
\includegraphics[width=2.2in]{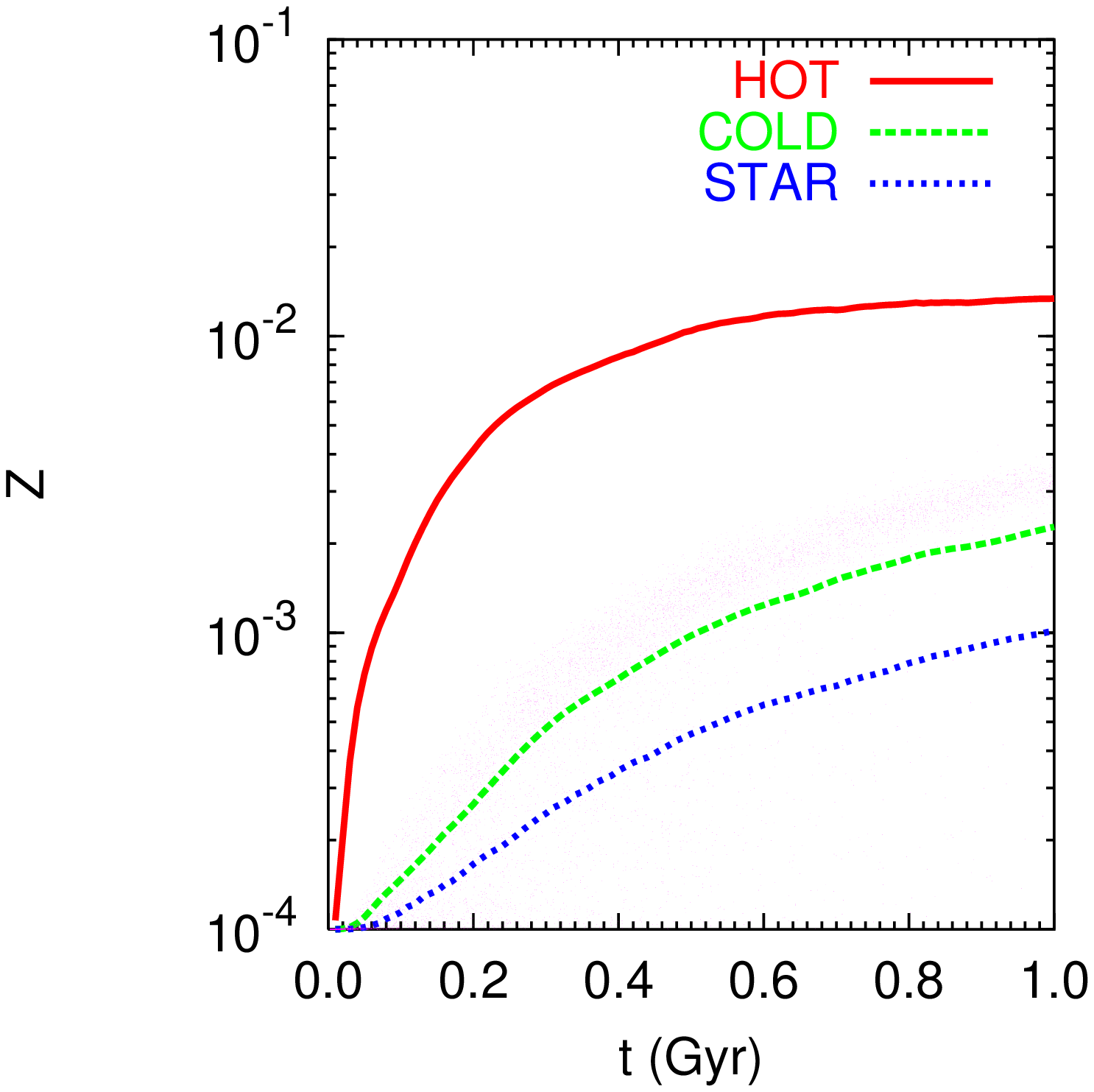} &
\includegraphics[width=2.2in]{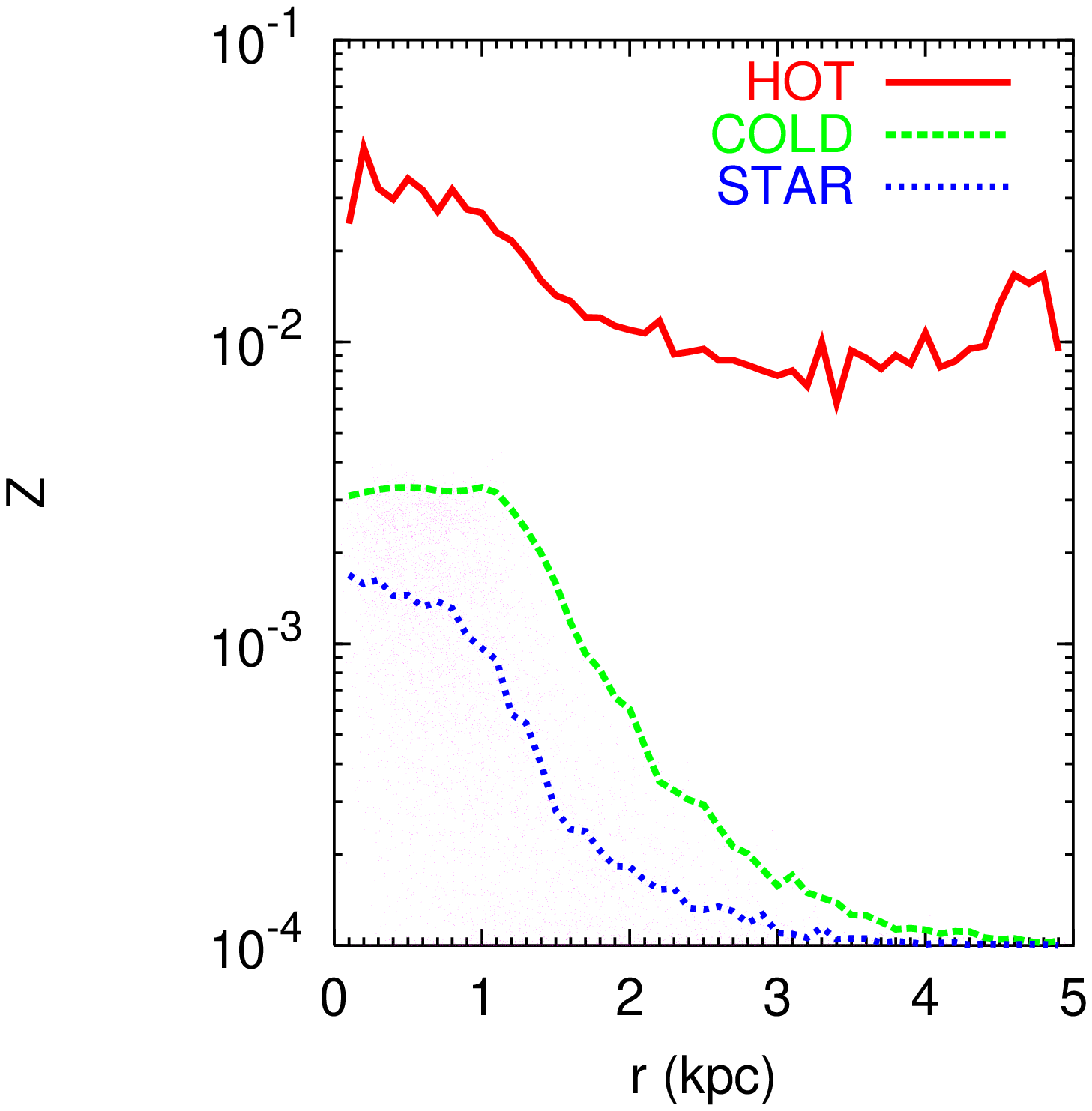}
\end{tabular}}
\caption{Temporal evolution of the metallicities (left) and their radial
distribution after 1 Gyr (right). Individual metallicities of newly born
stars are marked by dots.}
\label{z-t&z-r}
\end{figure}

The metal content of the diffuse gas and the clouds differs
significantly over the whole integration time
(Fig.~\ref{z-t&z-r}). Due to SNII and SNIa events the metallicity
of the hot phase exceeds that of the clouds by almost one order of
magnitude. The clouds mainly get their metals by condensation of
the hot phase. The central metallicity plateau (up to 1~kpc) of
the cold component is explained by the fact, that condensation is
not very efficient in that region, e.g.\ because of the central
``hole'' of the very hot diffuse ISM and a lack of metal enriched
material there. Moreover, the conditions in the center lead mainly
to evaporation of clouds which also prevents the mixing with the
metal enriched hot gas.

%%%%%%%%%%%%%%%%%%%%%%%%%%%%%%%%%%%%%%%%%%%%%%%%%%%%%%%%%%%%%%%%%%%%%%%%%%%%%

%---------------------------------------------------------------------------%
%---------------------------------------------------------------------------%

\vskip -0.5cm
\acknowledgements

The work was supported by the German Science Foundation (DFG) with the 
grants 436 UKR 18/2/99, 436 UKR 17/11/99 and the SFB 439 (University of 
Heidelberg). P.B. is grateful for the hospitality of the Astronomisches 
Rechen-Institut (Heidelberg), where the main part of this work has been 
done. The numerical models have been computed with the GRAPE-5 system at 
the Astronomical Data Analysis Center of the National Astronomical 
Observatory, Japan.

%%%%%%%%%%%%%%%%%%%%%%%%%%%%%%%%%%%%%%%%%%%%%%%%%%%%%%%%%%%%%%%%%%%%%%%%%%%%%

%%%%%%%%%%%%%%%%%%%%%%%%%%%%%%%%%%%%%%%%%%%%%%%%%%%%%%%%%%%%%%%%%%%%%%%%%%%%%

\end{article}

\vskip -0.5cm
\begin{thebibliography}{}

\bibitem[\protect\citeauthoryear{Berczik}{1999}]{Ber99}
Berczik P., 1999, A\&A, {\bf 348}, 371

\bibitem[\protect\citeauthoryear{Berczik}{2000}]{Ber2000}
Berczik P., 2000, Ap\&SS, {\bf 271}, 103

\bibitem[\protect\citeauthoryear{Burkert}{1995}]{B95}
Burkert A., 1995, ApJ, {\bf 447}, L25

\bibitem[\protect\citeauthoryear{Cowie et al.}{1981}]{CMcKO81}
Cowie L.L., McKee C.F. \& Ostriker J.P., 1981, ApJ, {\bf 247}, 908

\bibitem[\protect\citeauthoryear{K\"oppen et al.}{1998}]{KTH98}
K\"oppen J., Theis Ch. \& Hensler G., 1998, A\&A, {\bf 331}, 524

\bibitem[\protect\citeauthoryear{Mihos \& Hernquist}{1996}]{MH96}
Mihos J.C. \& Hernquist L., 1996, ApJ, {\bf 464}, 641

\bibitem[\protect\citeauthoryear{Miyamoto \& Nagai}{1975}]{MN75}
Miyamoto M. \& Nagai R., 1975, PASJ, {\bf 27}, 533

\bibitem[\protect\citeauthoryear{Navarro \& White}{1993}]{NW93}
Navarro J.F. \& White S.D.M., 1993, MNRAS, {\bf 265}, 271

\bibitem[\protect\citeauthoryear{Samland et al.}{1997}]{SHT97}
Samland M., Hensler G. \& Theis Ch., 1997, ApJ, {\bf 476}, 544

\bibitem[\protect\citeauthoryear{Theis et al.}{1992}]{TBH92}
Theis Ch., Burkert A. \& Hensler G., 1992, A\&A, {\bf 265}, 465

\bibitem[\protect\citeauthoryear{Theis \& Hensler}{1993}]{TH93}
Theis Ch. \& Hensler G., 1993, A\&A, {\bf 280}, 85

\end{thebibliography}
\end{document}